\documentclass[preprint,review,12pt]{elsarticle}

\usepackage{latexsym}
\usepackage{lipsum}
\usepackage{epsf}
\usepackage{float}
\usepackage{epsfig}
\usepackage{subfigure}
\usepackage{amsmath}
\usepackage{amssymb}
\usepackage{hyperref}
\usepackage{changes}
\usepackage{dcolumn}
\usepackage{bm}
\usepackage{xcolor}
\usepackage{tikz}
\usetikzlibrary{shapes}
\definecolor{blue}{RGB}{0,112,192}
\definecolor{lightblue}{RGB}{0,176,240}
\definecolor{green}{RGB}{0,176,80}
\definecolor{yellow}{RGB}{255,255,0}
\definecolor{orange}{RGB}{255,192,0}
\definecolor{red}{RGB}{255,0,0}
\definecolor{darkred}{RGB}{118,0,0}
\definecolor{purple}{RGB}{208,0,154}

\newcommand*{\tikzcirc}[2]{%
   \setbox0=\hbox{\strut}%
   \begin{tikzpicture}
     \useasboundingbox (-.25em,0) rectangle (.25em,\ht0);
     \filldraw[draw=#1,fill=#2] (0.0,0.35\ht0) circle[radius=.25em];
   \end{tikzpicture}%
}
\newcommand*{\tikzrect}[2]{%
   \setbox0=\hbox{\strut}%
   \begin{tikzpicture}
     \useasboundingbox (-.25em,0) rectangle (.25em,\ht0);
     \filldraw[draw=#1,fill=#2] (-0.25em,0.05em) rectangle (.25em,0.55em);
   \end{tikzpicture}%
}
\newcommand*{\tikzdiam}[2]{%
   \setbox0=\hbox{\strut}%
   \begin{tikzpicture}[rotate = 45]
     \useasboundingbox (-.25em,0) rectangle (.25em,\ht0);
     \filldraw[draw=#1,fill=#2] (-0.25em,0.0em) rectangle (.25em,0.5em);  
   \end{tikzpicture}%
}

\newcommand{\tikztri}[2]{\tikz{\node[draw=#1,fill=#2,isosceles triangle,isosceles triangle stretches,shape border rotate=90,minimum width=0.2cm,minimum height=0.2cm,inner sep=0pt] at (0,0) {};}}

\usepackage{lineno}
\journal{Powder Technology}

\begin{document}

\begin{frontmatter}



\title{Influence of Cross-section Shape on Granular Column Collapses\tnoteref{label1}}


\author[inst1]{Teng Man}

\affiliation[inst1]{organization={Key Laboratory of Coastal Environment and Resources of Zhejiang Province (KLaCER), School of Engineering, Westlake University},
            addressline={18 Shilongshan St.}, 
            city={Hangzhou},
            postcode={310024}, 
            state={Zhejiang},
            country={China}}

\author[inst2]{Herbert E. Huppert}
\affiliation[inst2]{organization={Institute of Theoretical Geophysics, King's College, University of Cambridge},
            addressline={King's Parade}, 
            city={Cambridge},
            postcode={CB2 1ST}, 
            state={},
            country={United Kingdom}}

\author[inst1]{Zaohui Zhang}

\author[inst1]{Sergio A. Galindo-Torres\corref{cor1}}
\tnotetext[label1]{\\}
\cortext[cor1]{Corresponding author,  Tel: +86 15158184891}
\ead{s.torres@westlake.edu.cn}

\date{\today}

\begin{abstract}
We investigate granular column collapses with different column cross-sections and associate the cross-section shape influence with a finite-size analysis. Previous research, confined to initially circular configurations, reviewed the importance of granular column collapse studies and concluded that the run-out distance scales with the initial aspect ratio. In this work, granular columns with three different types of initial column cross-sections (square, equilateral triangular, and rectangular cross-sections) are simulated using the discrete element method (DEM). We explore how non-circular cross-sections lead to different run-out distances. Based on the previously obtained finite-size analysis, we further link the initial radius in different directions to the relative size of a column and perform the finite-size analysis to explain the cross-section influence. In the end, a universal relationship, which includes frictional properties, relative sizes, and the cross-section influence, is proposed. Our results are shown to have direct relevance to various natural and engineering systems.
\end{abstract}




\end{frontmatter}


\section{Introduction}
\label{sec:intro}
Granular materials are ubiquitous in natural and engineering systems. The dynamics of granular column collapses play an important role and can bring insights into understanding the kinetics and rheology of complex granular systems in civil engineering, chemical engineering, pharmaceutical engineering, food processing, and geophysical flows \cite{Andreotti2013Granular}. In recent decades, progress has been made in terms of the rheological behavior of granular systems \cite{midi2004dense, trulsson2012transition, man2021granular}. Pouliquen et al. \cite{midi2004dense, pouliquen2006flow} concluded that the rheology of granular systems was controlled by inertial numbers ($\mu-I$ rheology), and continuum models based on $\mu-I$ rheology were proposed accordingly \cite{jop2006constitutive,Lagree2011granular,barker2015well}, which have made significant contributions to understanding and modeling the behavior of granular assemblies.

Initially, the behavior of granular column collapses was investigated to better understand the post-failure behavior of discrete systems, such as geophysical flows with similar flow mechanisms \cite{Roche2002ExperimentsOD,lube2004axisymmetric}. Roche et al. \cite{Roche2002ExperimentsOD,Roche2008ExperimentalOO} linked the dam-break experiments to the physics of pyroclastic flows, and argued that the behavior of granular flows is similar to pyroclastic flows when they are in certain regimes. Lube et al. \cite{lube2004axisymmetric,lube2005collapses} and Lajeunesse et al. \cite{lajeunesse2005granular} independently discovered that both the run-out distance and the final deposition height can be determined by the initial aspect ratio of the column. In particular, the normalized run-out distance, $\mathcal{R} = (R_\infty-R_i)/R_i$, where $R_\infty$ is the final radius of the granular pile and $R_i$ is the initial radius of the granular column, proportionally scales with the initial aspect ratio, $\alpha = H_i/R_i$, where $H_i$ is the initial height of the column, when $\alpha<\alpha_c$ (where $\alpha_c$ is a transition point determined by experimental results), and proportionally scales with $\alpha^{0.5}$ when $\alpha>\alpha_c$.

Zenit \cite{zenit2005computer} performed discrete element method (DEM) simulations on two dimensional (2D) granular column collapses, and confirmed that the shape of the final deposition was mainly determined by the initial aspect ratio. Staron and Hinch \cite{staron2005study,staron2007spreading} further investigated 2D granular collapses with DEM, and found that the inter-particle frictional coefficient played an important role in the run-out distance. Lacaze and Kerswell \cite{lacaze2009axisymmetric} studied axisymmetric granular collapses to test the viscoplasticity of granular materials. Lagr{\'e}e et al. \cite{Lagree2011granular} implemented the $\mu-I$ rheology to a Navier-Stokes solver to study the behavior of granular column collapses using a continuum approach. The results of the continuum approach with $\mu-I$ rheology showed good agreement with the results of DEM simulations. Farin et al. \cite{Farin2019RelationsBT} even associated the characteristics of granular column collapses with high-frequency seismic signals, and used it to further evaluate geological events. With experiments, Cabrerra et al. \cite{Cabrera2019GranularCC} and Warnett et al. \cite{warnett2014scalings} discovered that the relative size of the granular column could also influence the normalized run-out distance. They also found that, when the system size was large enough, the size effect can be neglected.

Based on previous studies, we implemented dimensional analysis, and obtained an effective aspect ratio, $\alpha_{\textrm{eff}} = \sqrt{1/(\mu_w +\beta\mu_p)}(H_i/R_i)$, where $\mu_w$ is the basal frictional coefficient, $\mu_p$ the inter-particle frictional coefficient, and $\beta=2.0$ a constant that can be physically interpreted as the ratio of contributions between inter-particle frictions and particle/boundary frictions \cite{man2020universality}. The effective aspect ratio includes the influence of both friction and the initial aspect ratio. Furthermore, we investigated the finite-size scaling of granular column collapses \cite{man2021finitesize}, and come up with a universal scaling equation to describe the run-out behavior. We showed that changing the system size could not only influence the normalized run-out distance but also shift the characteristic aspect ratio $\alpha_c$ that marks the transition of granular column collapses from a quasi-static regime to an inertial regime. A general scaling equation for the run-out distance of columns with different sizes was obtained \cite{man2021finitesize} and given by,
\begin{equation} \label{eq_fss}
    \begin{split}
        \mathcal{R} = \left({R_i}/{d}\right)^{-\beta_1/\nu}\mathcal{F}_r\left[(\alpha_{\textrm{eff}}-\alpha_{c\infty})\left({R_i}/{d}\right)^{1/\nu}\right]\ ,
    \end{split}
\end{equation}
where $\mathcal{F}_r[\cdot]$ is a scaling function, scaling parameters $\nu = 1.39\pm 0.14$ and $\beta_1 = 0.28\pm 0.04$ are obtained to best collapse all the data, $\alpha_{c\infty}$ is the transitional effective aspect ratio when the system size goes to infinity, and $d$ is the average particle diameter. The current $\mathcal{F}_{r}[\cdot]$ function still lacks a functional form, which limits the application of the finite-size scaling and the generalization of our findings to other granular flow cases. However, there is no contradiction between the finite-size solution and the power-law scaling. In fact, they are both related by the renormalization group formalism of statistical mechanics. The power-law relationship between $\mathcal{R}$ and $\alpha_{\textrm{eff}}$ (of systems with the same relative size) has a unique critical point, $\mathcal{R}_c$ and $\alpha_c$, which marks the transition from quasi-static collapses to inertial collapses. The finite-size solution with $R_i/d$ is utilized to, in some sense, collapse $\mathcal{R}$ and $\alpha_c$ for systems with different relative sizes. In other words, the power-law relationship is still embedded inside the $\mathcal{F}_{r}[\cdot]$ function, and should be recovered as the domain size diverges. We are also aware that it is of vital importance to obtain the exact function for $\mathcal{F}_{r}[\cdot]$, but the current results did not lead us to that stage, and we will continue to work on this problem.

So far, only granular column collapses of circular cylinders or 2D granular column collapses have been considered. We were somewhat surprised by our early toy experiments of the collapse of rectangular granular columns, where the final deposition approaches a circular pattern when the initial height is sufficiently large. Thus, we wondered how granular columns with different initial shapes collapse and what their final deposition patterns are. In this paper, we investigate the influence of cross-section shape on granular column collapses. Previous research focuses on pseudo-2D or axisymmetric granular column collapses, where the shape of the cross-section does not play a role in determining the run-out distance. However, non-axisymmetric cross-sections lead to different initial column radii in different directions. For instance, for a square cross-section, the initial radius in the diagonal direction is $\sqrt{2}$ times of that in the direction pointing from the cross-section center to the center of the edge. A different initial radius results in a different initial aspect ratio, which further leads to a different normalized run-out distance. This paper systematically explores the relationship between normalized run-out distance and effective aspect ratio in different directions resulting from different types of cross-sections, including square, rectangle , and equilateral triangle (We choose these types of cross-sections with right angles or sharp angles so that the direction of vertex and edge can be clearly and conveniently defined).  The paper is organized in the following way. In Section \ref{sec:simu}, the simulation setup and the associated numerical method are introduced. In Section \ref{sec:Result}, we present the simulation results, and further discuss the physical insight that follows the simulation results. We further implement the finite-size analysis that we obtained from our previous work \cite{man2021finitesize}. An experimental validation in Section \ref{sec:exp} will also be provided to show experimental evidence of the influence of cross-section shapes, before concluding remarks are provided in Section \ref{sec:conclude}. 


\section{Simulation method and setup}
\label{sec:simu}

\subsection{Discrete element method}

To explore the behavior of the collapse of granular columns, we implement a Voronoi-based sphero-polyhedral discrete element method (DEM) \cite{cundall1979discrete,galindo2010molecular}\added{,} so that we could obtain detailed particle-scale information during \deleted{a }column collapse\added{s}. We note that the particle shapes could significantly affect the deposition morphology but, in this study, we focus on using Voronoi-based particles to investigate a general behavior similar to that of sand particles.

The sphero-polyhedra method was initially introduced by Pourning \cite{pournin2005generalization} for the simulation of complex-shaped DEM particles. Later, it was modified by Alonso Marroquin \cite{alonso2008spheropolygons}, who introduced a multi-contact approach in 2D allowing the modelling of non-convex shapes and was extended to 3D by Galindo-Torres et. al. \cite{galindo2010molecular}. A sphero-polyhedron is a polyhedron that has been eroded and then dilated by a sphere. The result is a polyhedron of similar dimensions but with rounded corners.


An advantage of the sphero-polyhedra technique is that it allows for an easy and efficient definition of contact detection and force calculation between particles. This is due to the smoothing of edges of all geometric features by circles (in 2D) or spheres (in 3D). A particle is defined as a polyhedron, i.e. a set of vertices, edges and faces, where each one of these geometrical feature is dilated by a sphere.

Since two contacting particles are dilated by their sphero-radii $R_1$ and $R_2$, there exists a contact when the distance between two geometric features is less than the addition of their corresponding dilating radii, and the corresponding contact overlap $\delta$ can be calculated accordingly. The advantage of the sphero-polyhedra technique becomes evident since this definition is similar to the contact force calculation of two spheres\cite{belheine2009numerical}. In our simulation, three types of contacts (vertex-vertex contact, edge-edge contact, and vertex-face contact) are considered. For these types of contacts, we implement a Hookean contact model with energy dissipation to calculate the interactions between particles as illustrated in Ref. \cite{man2020universality}. At each time step, the overlap and the tangential relative displacement between adjacent particles, $\delta$, are checked, and the normal and tangential contact forces can be calculated accordingly.

In this study, since we use Voronoi-based particles, no rolling resistance need be considered. The motion of particles is then calculated by step-wise resolution of Newton's second law with the normal and contact forces mentioned before. The same neighbor detection and force calculation algorithms have already been discussed and validated in previous studies. This DEM formulation has been validated before with experimental data \cite{cabrejos-hurtado2016,belheine2009numerical} and is included in the MechSys open source multi-physics simulation library \cite{galindo2013coupled}.

\subsection{Simulation setup}

\begin{figure}[!ht]
  \centering
  \includegraphics[scale = 0.35]{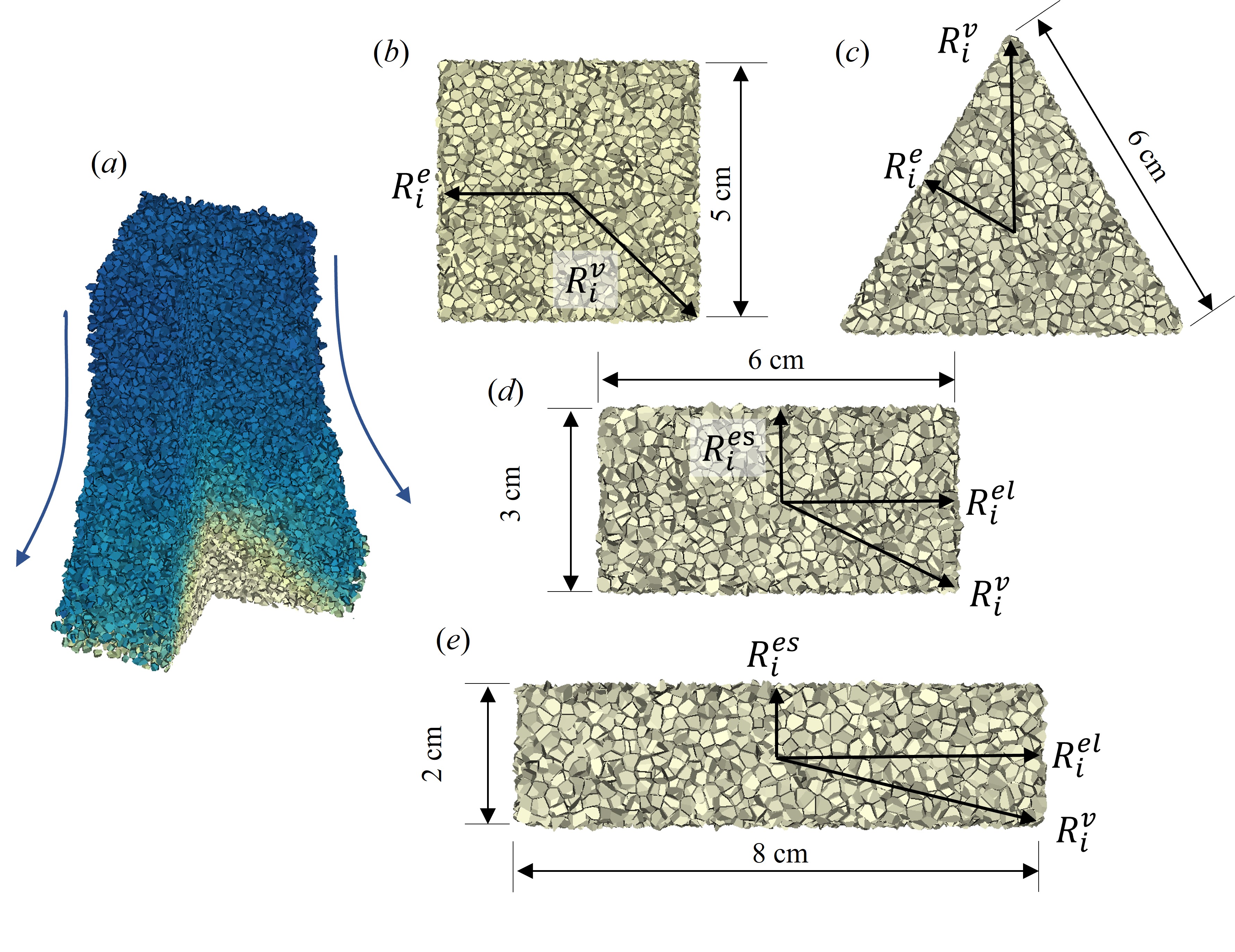}
  \caption{(a) shows the collapse of a granular column with square cross-section. Part of particles are cut from the figure to reveal the un-moving core during the collapse. (b)-(e) show four types of cross-sections we use in this work.}
  \label{fig-setup}
\end{figure}

The DEM in this work has already been validated in previous studies to simulate the behavior of granular materials in various conditions, such as triaxial tests of granular soils \cite{galindo2010molecular}, contact erosion phenomena \cite{GalindoTorres2015AMA}, and axisymmetric dry granular column collapses \cite{man2020universality,man2021finitesize}. We implement this DEM model to granular column collapses with different initial cross-sections. As shown in Fig. \ref{fig-setup}, we create granular columns with three different types of initial cross-sections: 1) columns with square cross-sections, where the side length $L_s = 5$ cm; 2) columns with equilateral triangular cross-sections, where the side length $L_s = 6$ cm; 3) columns with rectangular cross-section with side lengths 6 cm $\times$ 3 cm and 8 cm $\times$ 2 cm. The cross-section size of granular columns are chosen so that the edge length is approximately more than 10 times the particle size, and the area of the cross-section is around 20 cm$^2$. We exclude simulations with larger cross-section sizes because of the correspondingly unacceptable computational time. For each type of granular column, we vary the inter-particle frictional coefficient ($\mu_p = 0.1, 0.2, 0.4$), while keeping the particle-boundary frictional coefficient constant at $\mu_w = 0.4$. For simulations with square and triangular cross-sections, we also implement $\mu_p = 0.6$. The height of granular columns  varies from 1 cm to 50 cm to obtain various initial aspect ratios. 

The original sphero-polyhedral granular packing is established using the 3D Voronoi scheme with the \textit{Voro}++ package \cite{Rycroft2009VOROAT}. The average particle size is $d=0.2$ cm, so that the size of particles is similar to the size of typical medium sized river sand and, at the same time, small enough compared to the length of the cross-section of granular columns. Then 20\% of particles are randomly chosen and deleted to create an initial packing with a solid fraction $\phi_s$ of 0.8 to keep the initial solid fraction the same as our previous work \cite{man2020universality}. The initial solid fraction of 80\% is, on one hand, for reducing the volume change during the collapse process, on the other hand, for introducing randomness. We acknowledge the influence of the initial solid fraction on the collapse and deposition of granular columns, and will conduct further studies to address such influence in the future. Besides, we should be awared that, due to the property of 3D Voronoi structure, the initial packing is more similar to a fractured porous rock than to loosely packed sand.

Then, we remove the container in the simulation, release particles, and let them flow under gravity. During the collapse of a granular column, the stored potential energy will be transformed into kinetic energy and dissipated through particle collisions. In the end, a stable granular pile can be obtained with a final run-out distance, $R_{\infty}(\theta)$, where $\theta$ is the direction angle in the horizontal plane showing that, for a granular column with non-circular cross-section, the final run-out distance varies with respect to the direction in which we take the measurement. In detail, for columns with square cross-sections, we present measurements in the eight directions of 0$^{\circ}$, 45$^{\circ}$, 90$^{\circ}$, 135$^{\circ}$, 180$^{\circ}$, 225$^{\circ}$, 270$^{\circ}$, and 315$^{\circ}$. For columns with equilateral triangular cross-sections, we present measurements in the six directions of 30$^{\circ}$, 90$^{\circ}$, 150$^{\circ}$, 210$^{\circ}$, 270$^{\circ}$, and 330$^{\circ}$. For columns with rectangular cross-section of length $6$ cm and width $3$ cm, we present measurements in the eight directions of 0$^{\circ}$, 26.6$^{\circ}$, 90$^{\circ}$, 153.4$^{\circ}$, 180$^{\circ}$, 206.6$^{\circ}$, 270$^{\circ}$, and 333.4$^{\circ}$. Similarly, for columns with rectangular cross-section of length $8$ cm and width $2$ cm, we present the measurement in the eight directions of 0$^{\circ}$, 14$^{\circ}$, 90$^{\circ}$, 166$^{\circ}$, 180$^{\circ}$, 194$^{\circ}$, 270$^{\circ}$, and 346$^{\circ}$. These measuring directions can be classified into two groups: i) edge directions (for rectangular cross-section, the edge direction can be further classified into longitudinal and width  edge directions); and ii) vertex directions.

\section{Results and discussions}
\label{sec:Result}



\subsection{Granular columns with square cross-sections}

\begin{figure}[!ht]
  \centering
  \includegraphics[scale = 0.35]{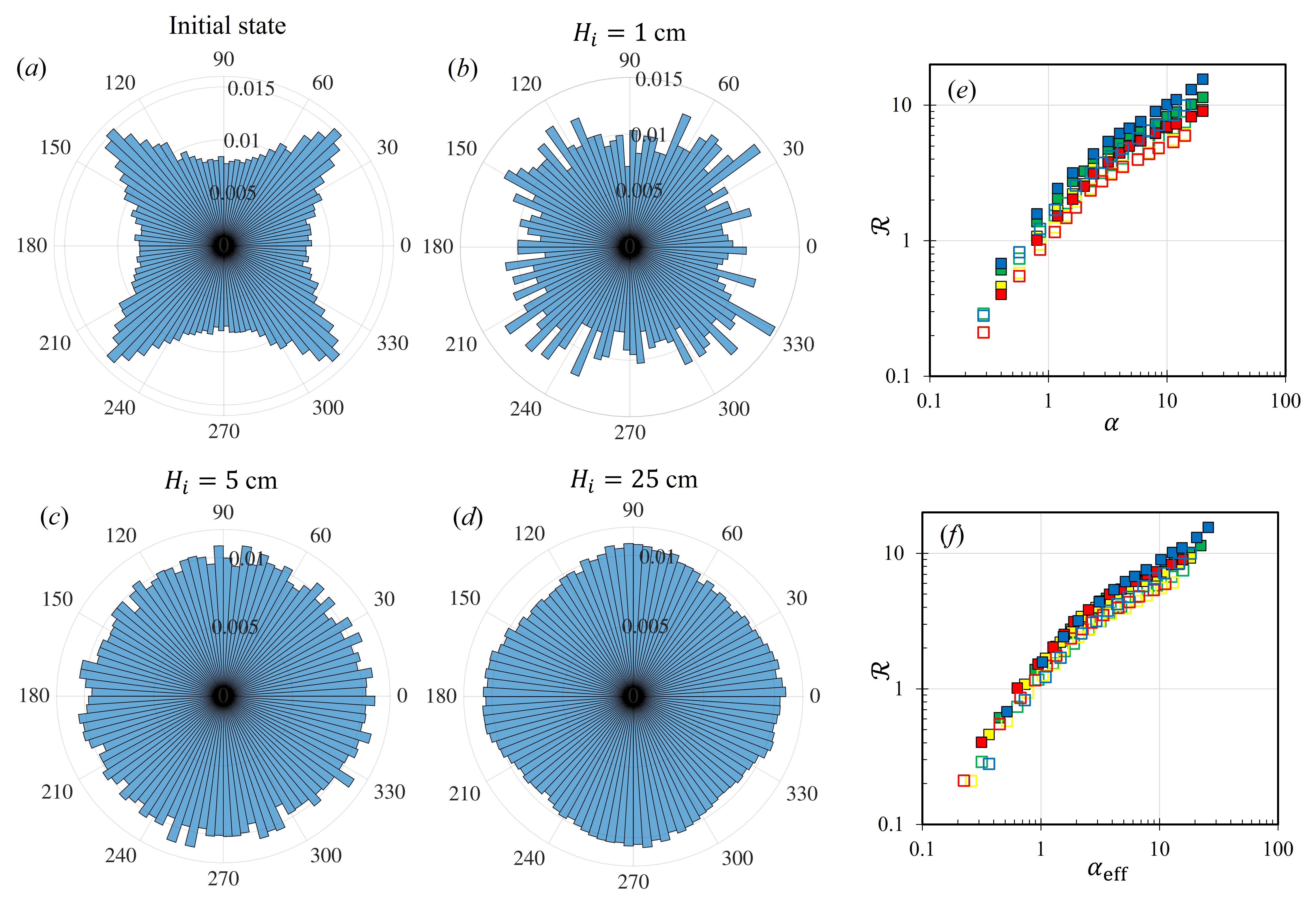}
  \caption{(a)-(d) show the polar distributions of particle numbers in different directions for granular column collapses with square cross-sections and faces parallel to the edge of the paper. The numbers on the vertical axis represent the percentage of particles locating in certain directions, and the numbers on the circumferential axis is the direction angle measured counter-clockwise. (e) denotes the relationship between the normalized run-out distance, $\mathcal{R}$, and the initial aspect ratio, $\alpha$, where markers represent simulations $\mu_p = 0.1$ in the edge direction \protect\tikzrect{black}{blue}, $\mu_p = 0.2$ in the edge direction \protect\tikzrect{black}{green}, $\mu_p = 0.4$ in the edge direction \protect\tikzrect{black}{yellow}, $\mu_p = 0.6$ in the edge direction \protect\tikzrect{black}{red}, $\mu_p = 0.1$ in the vertex direction \protect\tikzrect{blue}{white}, $\mu_p = 0.2$ in the vertex direction \protect\tikzrect{green}{white}, $\mu_p = 0.4$ in the vertex direction \protect\tikzrect{yellow}{white}, $\mu_p = 0.6$ in the vertex direction \protect\tikzrect{red}{white}. (f) plots $\mathcal{R}$ against the effective aspect ratio, $\alpha_{\textrm{eff}}$, with the same markers as those in Fig. (e).}
  \label{fig-square}
\end{figure}

For granular columns with square cross-sections, we take measurements of initial radius and final deposition radius in both edge and vertex directions. The initial radius in the edge direction is $R_i^e =$ 2.5 cm, and the final deposition radius in the edge direction is denoted as $R_{\infty}^e$. Similarly, the initial radius in the vertex direction is $R_i^v = \sqrt{2}R_i^e$, and the final deposition radius in the vertex direction is denoted as $R_{\infty}^v$. Thus, we obtain the following key parameters of initial aspect ratios and effective aspect ratios,
\begin{linenomath}
\begin{subequations}
\begin{align}
\alpha^e &= H_i/R_i^e,\ \ \alpha^v = H_i/R_i^v, \\
\alpha_{\textrm{eff}}^e &= \alpha^e\sqrt{1/(\mu_w + \beta\mu_p)}, \ \alpha_{\textrm{eff}}^v = \alpha^v\sqrt{1/(\mu_w + \beta\mu_p)},
\end{align}
\end{subequations}
\end{linenomath}
where $\alpha^e$, $\alpha^v$ are the initial aspect ratios in the edge and vertex directions, $\alpha_{\textrm{eff}}^e$ and $\alpha_{\textrm{eff}}^v$ are the effective aspect ratios in both edge and vertex directions, $\mu_w$ is the frictional coefficient between particles and the horizontal plane, $\mu_p$ is the inter-particle frictional coefficient, and $\beta = 2.0$ as suggested by Ref. \cite{man2020universality}. We could also obtain the normalized run-out distances in both directions, $\mathcal{R}^e$ and $\mathcal{R}^v$, as
\begin{linenomath}
\begin{subequations}
\begin{align}
\mathcal{R}^e &= (R_{\infty}^e - R_i^e)/R_i^e, \\
\mathcal{R}^v &= (R_{\infty}^v - R_i^v)/R_i^v.
\end{align}
\end{subequations}
\end{linenomath}

The choice of only measuring run-out behaviors in both edge and vertex directions for simplifying the analysis. In a more ideal condition, we can measure the run-out distance in any arbitrary directions so that our analysis can be more complete. In Fig. \ref{fig-square}(a)-(d), we plot, for a range of initial heights, the polar histogram of the number of particles in different directions to show the initial and final pattern of the granular system. To plot the polar histogram, we divide the round angle (360$^{\circ}$) into 100 pieces so that d$\theta=3.6^{\circ}$. Each bar in the polar histogram is obtained by counting the number of particles presenting within $(\theta, \theta+\textrm{d}\theta]$, and dividing it by the total number of particles in the system. Fig. \ref{fig-square}(a) shows the initial state of the system, where the majority of particles locate in the vertex direction, since the initial radius of it is larger than that of the edge direction. For systems with $H_i =$ 1 cm [Fig. \ref{fig-square}(b)], the polar histogram still shows features of squares, where more particles locate in the vertex direction. As we further increase the initial height of the granular column, the final deposition becomes approximately circular [Fig. \ref{fig-square}(c) where $H_{i} = 5$ cm] with an increased horizontal length scale. This indicates that, as we increase the initial aspect ratio of the granular column, the system evolves from a quasi-static system\cite{man2020universality}, where most particles remain stationary during the collapse, to an inertial system, where most particles participate in the collapse event, and, during the evolution from a quasi-static system to an inertial collapsing system, memory of its initial state is forgotten; hence, a circular deposition pattern is developed. We note that, to be consistent with Ref. \cite{man2020universality}, the quasi-static collapse is for describing the overall collapse and deposition behaviors, and does not indicate that no particle inertia is playing a role during the early stage of granular column collapses. Further increasing the initial aspect ratio leads to more drastic changes to the deposition pattern, where particles in the edge direction propagate much further than particles in the vertex direction, as shown in Fig. \ref{fig-square}(d) where $H_i =$ 25 cm. This usually happens in the liquid-like collapse of granular columns in the phase diagram in Ref. \cite{man2020universality}. We obtain the general conclusion that, for a granular column with square cross-section, the run-out distance in the edge direction is larger than that in the vertex direction, especially when the initial aspect ratio is large enough to trigger a liquid-like collapse. This is because the initial aspect ratio in the edge direction is effectively larger than that in the vertex direction for the same granular column.

We further explore the run-out distance in different directions in Fig. \ref{fig-square}(e) and (f), where we plot the relationship between $\mathcal{R}$ and $\alpha$, and the relationship between $\mathcal{R}$ and $\alpha_{\textrm{eff}}$. In Fig. \ref{fig-square}(e), we show that the normalized run-out distance in both directions has a similar behavior with increasing the initial aspect ratio. For granular columns with the same initial aspect ratio but different inter-particle frictional coefficients, a smaller inter-particle frictional coefficient results in a larger run-out distance, which is the same for both edge and vertex directions. Interestingly, the normalized run-out distance in an edge direction is larger than that in a vertex direction, even when the initial aspect ratio is the same. 

We further investigate the relationship between $\mathcal{R}$ and $\alpha_{\textrm{eff}}$, and expect that changing the $x-$axis from $\alpha$ to $\alpha_{\textrm{eff}}$ could lead to the collapse of normalized run-out distance data of simulations with different frictional coefficients, according to our previously published work \cite{man2020universality}. Fig. \ref{fig-square}(f) shows that, for simulation results in either an edge or vertex direction, all the simulation data nicely collapse onto one curve. However, the data of measurements in the vertex direction are still smaller than those of measurements in the edge direction. This is surprising, especially after we analyzed the finite-size scaling of granular column collapses, where larger system sizes (larger $R_i/d$) lead to larger normalized run-out distance when the effective aspect ratios, $\alpha_{\textrm{eff}}$, are same \cite{man2021finitesize}. The finite-size scaling of granular column collapses implies that, with the same $\alpha_{\textrm{eff}}$, the $\mathcal{R}$ in the vertex direction should be larger than that in the edge direction. This paper aims to solve such a counter-intuitive situation (in Sections \ref{sec-equivR} and \ref{sec-fss}).

\subsection{Granular columns with equilateral triangular cross-sections}

\begin{figure}[!ht]
  \centering
  \includegraphics[scale = 0.35]{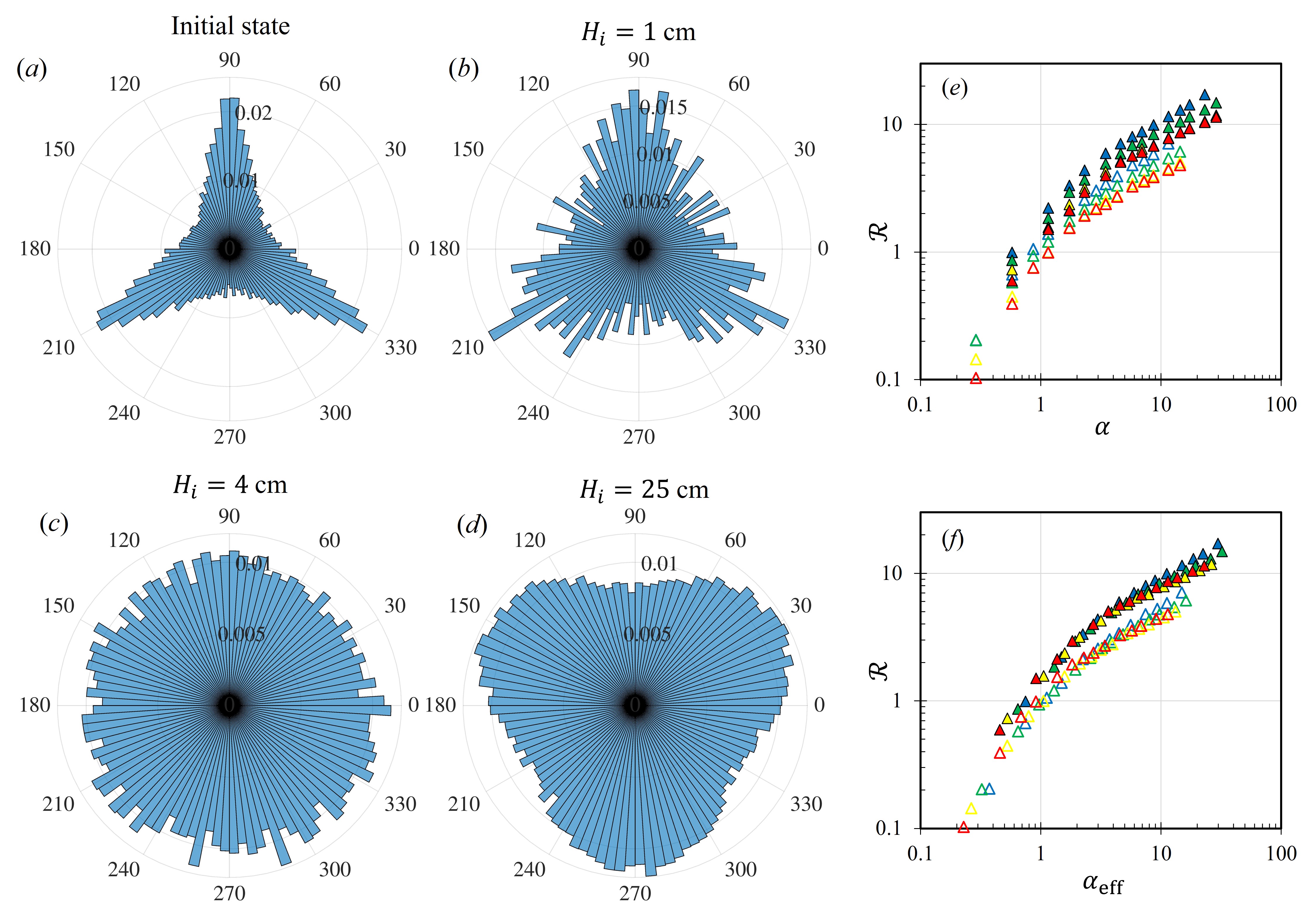}
  \caption{(a)-(d) show the polar distributions of particle numbers in different directions for granular column collapses with equilateral triangular cross-sections with a vertex at the top of the page. (e) and (f) denote the relationship between $\mathcal{R}$ and $\alpha$, and the relationship between $\mathcal{R}$ and $\alpha_{\textrm{eff}}$, where markers represent simulations with $\mu_p = 0.1$ in the edge direction \protect\tikztri{black}{blue}, $\mu_p = 0.2$ in the edge direction \protect\tikztri{black}{green}, $\mu_p = 0.4$ in the edge direction \protect\tikztri{black}{yellow}, $\mu_p = 0.6$ in the edge direction \protect\tikztri{black}{red}, $\mu_p = 0.1$ in the vertex direction \protect\tikztri{blue}{white}, $\mu_p = 0.2$ in the vertex direction \protect\tikztri{green}{white}, $\mu_p = 0.4$ in the vertex direction \protect\tikztri{yellow}{white}, $\mu_p = 0.6$ in the vertex direction \protect\tikztri{red}{white}.}
  \label{fig-triangle}
\end{figure}

We investigate the collapse of granular columns with equilateral triangular cross-section. Again, to simplify the analysis, we only take measurement in edge and vertex direction. Such a cross-section has three edge directions and three vertex directions, so we can define $\alpha^e$, $\alpha^v$, $\alpha_{\textrm{eff}}^e$, $\alpha_{\textrm{eff}}^v$, $\mathcal{R}^e$, and $\mathcal{R}^v$ as we did for columns with a square cross-section. In Fig. \ref{fig-triangle}(a)-(d), we plot polar histograms of the initial state and final depositions of columns with different initial heights. The polar distributions of the initial state shows that initially most particles locate in the vertex direction. Similar to columns with square cross-sections, when the initial height is relatively small [Fig. \ref{fig-triangle}(b)], the final pattern is still somewhat triangular. Increasing the initial height leads to changing the deposition pattern from approximately a triangle to approximately a circle [Fig. \ref{fig-triangle}(c)], which implies the memory of its initial triangular cross-section has been largely forgotten during the collapse. Fig. \ref{fig-triangle}(d) shows that, when the initial height of a granular column is large enough, the final deposition pattern becomes no longer circular. The run-out distance in the edge direction is much larger than that in the vertex direction, which results in a reverse of the equilateral triangle so that the original edge directions become vertex directions in the final deposition. It looks as if the triangle was flipped over. This surprising, but important, behavior is exactly the same as that when the cross-section is a square.

We plot the relationship between $\mathcal{R}^e$ (or $\mathcal{R}^v$) and $\alpha^e$ (or $\alpha^v$), as well as between $\mathcal{R}^e$ (or $\mathcal{R}^v$) and $\alpha_{\textrm{eff}}^e$ (or $\alpha_{\textrm{eff}}^v$) in Fig. \ref{fig-triangle}(e) and (f). Similar to the case for a axisymmetric granular column collapse, either $\mathcal{R}^e$ or $\mathcal{R}^v$ approximately scales proportional to $\alpha^e$ or $\alpha^v$ when the aspect ratio is less than a threshold, or scales proportional to $(\alpha^e)^{0.5}$ or $(\alpha^v)^{0.5}$ when the aspect ratio is larger than that threshold. Meanwhile, the normalized run-out distance is influenced by the frictional coefficient, as we predicted in our previous work \cite{man2020universality}. Decreasing the inter-particle frictional coefficient can not only change the $\mathcal{R}-\alpha$ curve but also change the critical initial aspect ratio. Additionally, for a same initial aspect ratio, $\mathcal{R}^e$ is apparently larger than $\mathcal{R}^v$. We shift the $x-$axis to $\alpha_{\textrm{eff}}$ in Fig. \ref{fig-triangle}(f), and find that the behavior is similar to that of square granular columns. The gap between measurements in the edge direction and in the vertex direction is even larger than that when the cross-section is square shaped. This is explained by the fact that for a square cross-section, $R_i^v/R_i^e$ is only 1.414, while the initial radius ratio is 2.0 for a triangular cross-section.

\subsection{Granular columns with rectangular cross-sections}

\begin{figure}[!ht]
  \centering
  \includegraphics[scale = 0.35]{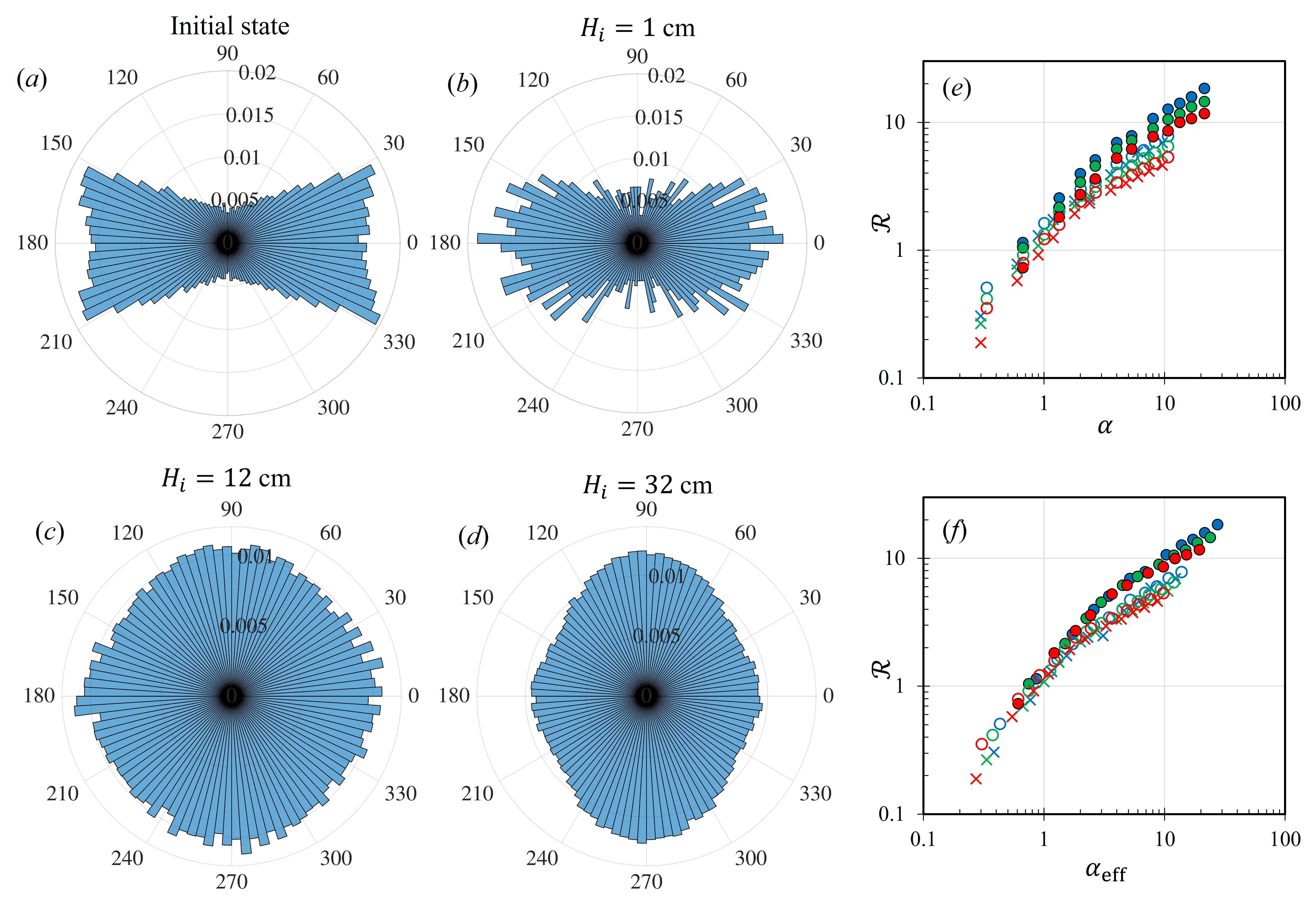}
  \caption{(a)-(d) show the polar distributions of particle numbers in different directions for granular column collapses with an initial $6\times3$ rectangular cross-section. (e) and (f) denote the relationship between $\mathcal{R}$ and $\alpha$, and the relationship between $\mathcal{R}$ and $\alpha_{\textrm{eff}}$, where markers represent simulations with $\mu_p = 0.1$ in the short-edge direction \protect\tikzcirc{black}{blue}, $\mu_p = 0.2$ in the short-edge direction \protect\tikzcirc{black}{green}, $\mu_p = 0.4$ in the short-edge direction \protect\tikzcirc{black}{red}, $\mu_p = 0.1$ in the long-edge direction \protect\tikzcirc{blue}{white}, $\mu_p = 0.2$ in the long-edge direction \protect\tikzcirc{green}{white}, $\mu_p = 0.4$ in the long-edge direction \protect\tikzcirc{red}{white}, $\mu_p = 0.1$ in the vertex direction \textcolor{blue}{$\times$}, $\mu_p = 0.2$ in the vertex direction \textcolor{green}{$\times$}, and $\mu_p = 0.4$ in the vertex direction \textcolor{red}{$\times$}.}
  \label{fig-rect1}
\end{figure}

Another set of simulations we present here is the collapse of granular columns with two different rectangular cross-sections ($6 \times 3$ and $8 \times 2$), where we take measurements in three directions: (1) the vertex direction, where its initial radius is $R_i^v$; (2) the long-edge direction, where its initial radius is $R_i^{el}$; and (3) the short-edge direction, where its initial radius is $R_i^{es}$. In this study, the long-edge direction is the direction parallel to the longer edge of the cross-section, and the short-edge direction is the one parallel to the short edge of the cross-section. For granular columns with a $6 \times 3$ rectangular cross-sections, $R_i^{el} = 3$ cm, $R_i^{es} = 1.5$ cm, and $R_i^v \approx 3.35$ cm. Similarly, for the granular columns with an $8 \times 2$ rectangular cross-section, $R_i^{el} = 4$ cm, $R_i^{es} = 1$ cm, and $R_i^v \approx 4.12$. We notice that the difference between $R_i^{el}$ and $R_i^v$ is small, so we expect that the run-out distance in these two directions will not differ much from one another. Also, since the initial aspect ratio in the short-edge direction is always the largest among three directions, we expect to obtain a larger run-out distances in this direction.

\begin{figure}[!ht]
  \centering
  \includegraphics[scale = 0.35]{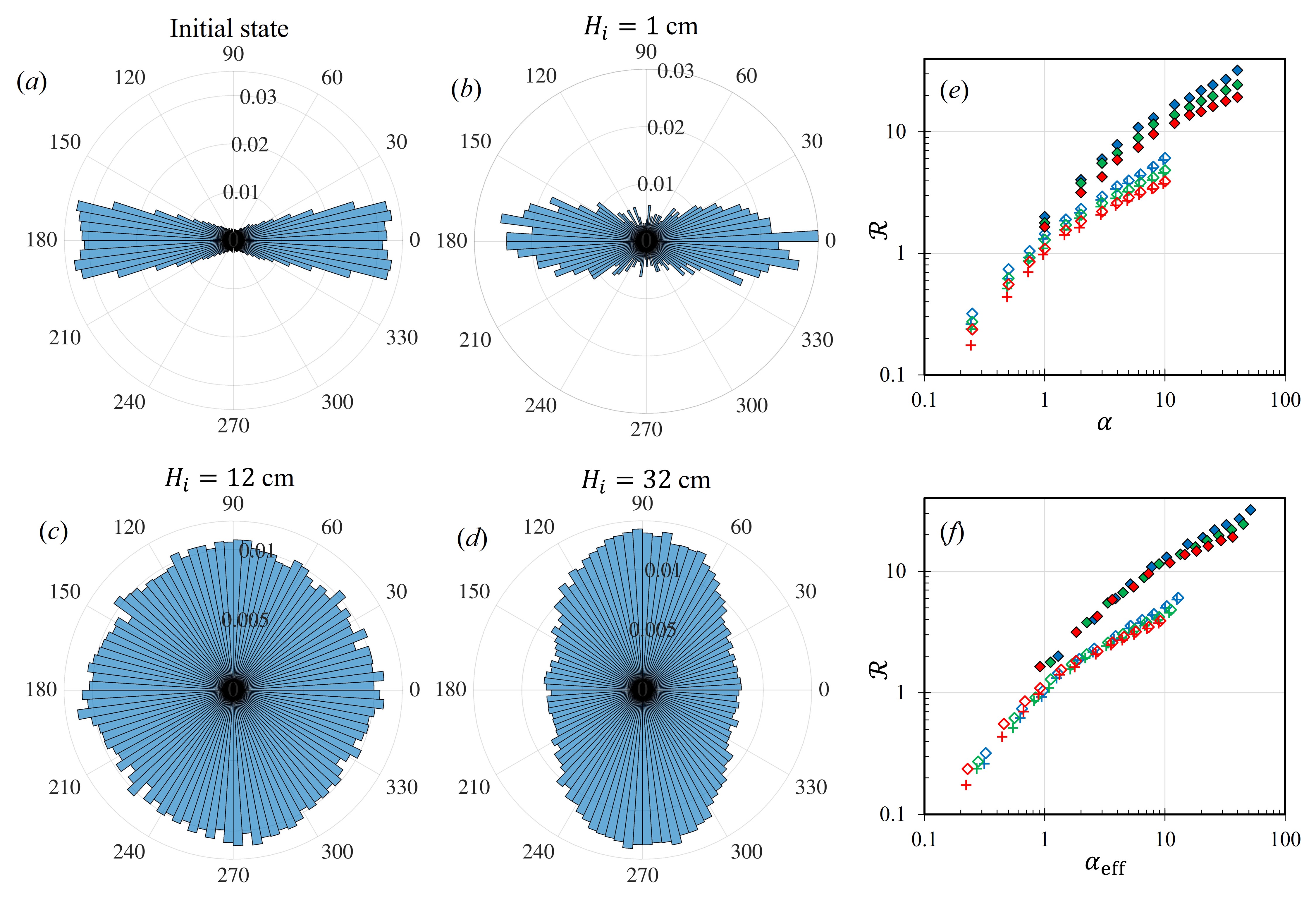}
  \caption{(a)-(d) show the polar distributions of particle numbers in different directions for granular column collapses with an initial $8\times2$ rectangular cross-sections. (e) and (f) denote the relationship between $\mathcal{R}$ and $\alpha$, and the relationship between $\mathcal{R}$ and $\alpha_{\textrm{eff}}$, where markers represent simulations with $\mu_p = 0.1$ in the short-edge direction \protect\tikzdiam{black}{blue}, $\mu_p = 0.2$ in the short-edge direction \protect\tikzdiam{black}{green}, $\mu_p = 0.4$ in the short-edge direction \protect\tikzdiam{black}{red}, $\mu_p = 0.1$ in the long-edge direction \protect\tikzdiam{blue}{white}, $\mu_p = 0.2$ in the long-edge direction \protect\tikzdiam{green}{white}, $\mu_p = 0.4$ in the long-edge direction \protect\tikzdiam{red}{white}, $\mu_p = 0.1$ in the vertex direction \textcolor{blue}{+}, $\mu_p = 0.2$ in the vertex direction \textcolor{green}{+}, and $\mu_p = 0.4$ in the vertex direction \textcolor{red}{+}.}
  \label{fig-rect2}
\end{figure}

We show the initial polar distribution of particles in Fig. \ref{fig-rect1}(a) and Fig. \ref{fig-rect2}(a). In the initial configuration, most particles locate in the vertex and long-edge directions The pattern varies when we change the initial height of the column. When the initial height is short, the final deposition pattern approximately remains rectangular [Fig. \ref{fig-rect1}(b) and Fig. \ref{fig-rect2}(b)]. Surprisingly, when $H_i \approx 12$ for $6 \times 3$ rectangular columns and $8 \times 2$ rectangular columns, the final deposition pattern already becomes circular, which indicates that the increase of the run-out distance in the short-edge direction is much stronger than the other two directions [Fig. \ref{fig-rect1}(c) and Fig. \ref{fig-rect2}(c)]. With further increase of the initial height of the column, we observe that the particles in the short-edge direction out-run those in the other two directions since the initial aspect ratio in the short-edge direction is larger than that in both the long-edge and the vertex direction, as can be seen in Fig. \ref{fig-rect1}(d) and Fig. \ref{fig-rect2}(d).

\begin{figure}[!ht]
  \centering
  \includegraphics[scale = 0.4]{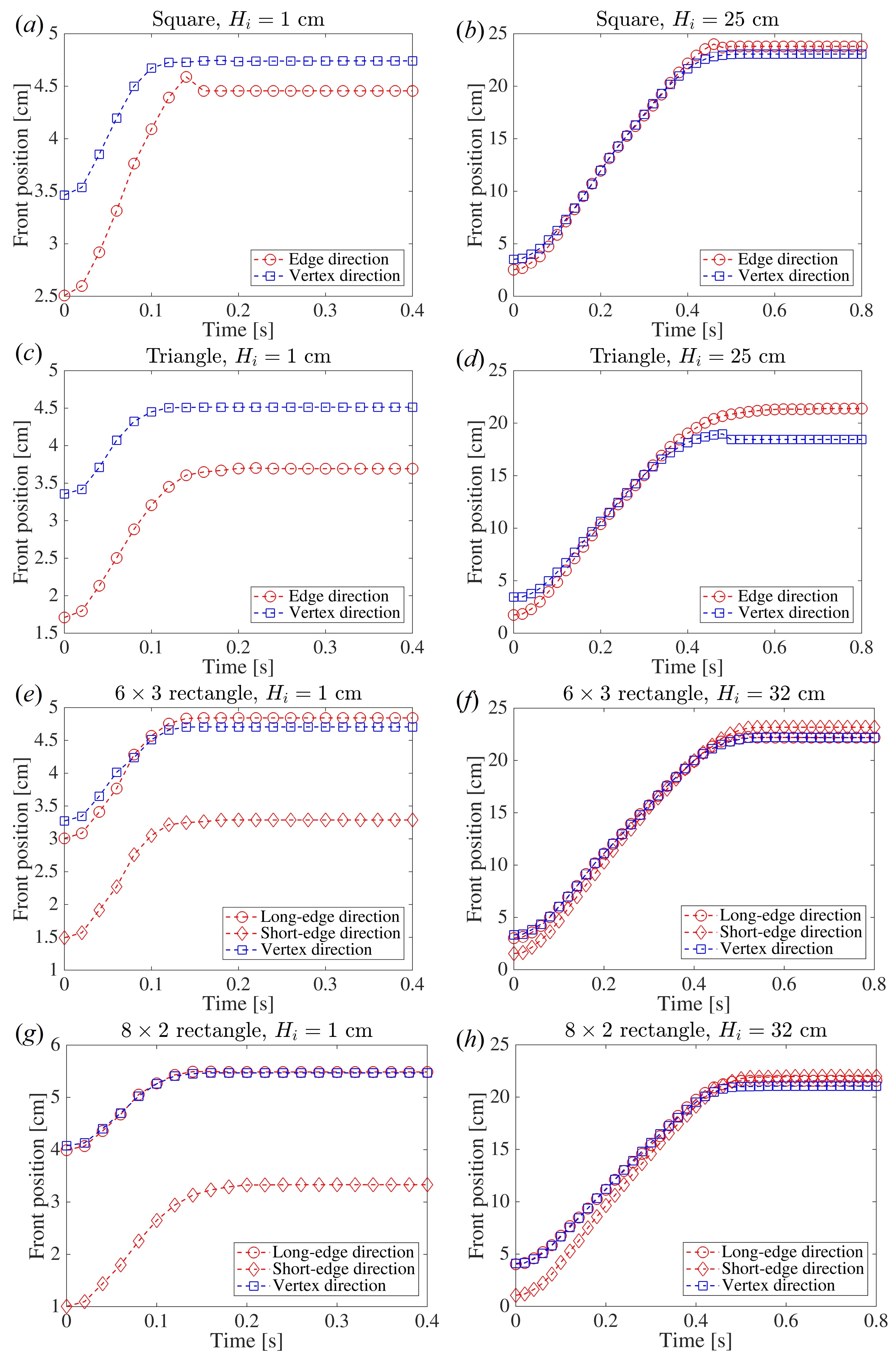}
  \caption{Relationship between the front position in different directions and time of granular columns with different initial heights and different types of cross-sections.}
  \label{fig-runouttime}
\end{figure}

What we have observed can also be validated by the plot of the $\mathcal{R}$-$\alpha$ relationship and the plot of the $\mathcal{R}$-$\alpha_{\textrm{eff}}$ relationship in Figs. \ref{fig-rect1}(e)-(f) and Figs. \ref{fig-rect2}(e)-(f). On one hand, since the initial radius in the vertex direction is similar to that in the long-edge direction, their results of normalized run-out distances are also similar. One the other hand, since the initial radius of the short-edge direction is much smaller than that of the other two directions, we see that the $\mathcal{R}^{es}$ - $\alpha^{es}$ relationships are always above the $\mathcal{R}^{el}$ - $\alpha^{el}$ relationships and $\mathcal{R}^v$ - $\alpha^v$ relationships, and it is the same for the $\mathcal{R}^{es}$ - $\alpha_{\textrm{eff}}^{es}$ relationships.

Failing to collapse all the data with even the effective aspect ratios leads us to re-think the way we normalize the run-out distances. Particles are connected with each other with inter-particle collisions, which form force networks. Mehta et al. \cite{mehta2007granular} and our previous work \cite{man2021finitesize} suggested that particles tend to move collectively, especially when the initial height is large enough to generate strong inertia among particles. Non-local effect tend to be strong, so that using the initial radius in a direction to normalize the run-out distance in that direction might not be an optimal choice.

We plot the relationship between the front position in different directions and the collapse time in Fig. \ref{fig-runouttime}. The left column of the figure shows the collapse of granular columns with a small initial height, while the right column plots those with a large initial height. For short columns, a larger initial radius often leads to a larger final front position. However, we can observe that the front position in the edge direction (or short-edge direction for rectangular cross-sections) is catching up with the front positions in other directions. In Fig. \ref{fig-runouttime}(e), we find that particles in the long-edge direction, while lagging behind those in the vertex direction at the beginning, quickly overrun particles in other directions even with such a small initial height. This shows clearly that particles in the edge direction have a better chance to acquire a larger run-out distance. This phenomenon is not obvious when the cross-section is an $8\times2$ rectangle. That is because, when the cross-section is an $8\times2$ rectangle, the difference between $R_i^{v}$ and $R_i^{el}$ is almost negligible. Figs. \ref{fig-runouttime}(b), (d), (f), and (h) also confirm what we have seen in Figs. \ref{fig-square}, \ref{fig-triangle}, \ref{fig-rect1}, and \ref{fig-rect2} that, at the end of the collapse, the front position in the edge direction (or in the short-edge direction) is always larger than that in other directions. 

\begin{figure}[!ht]
  \centering
  \includegraphics[scale = 0.3]{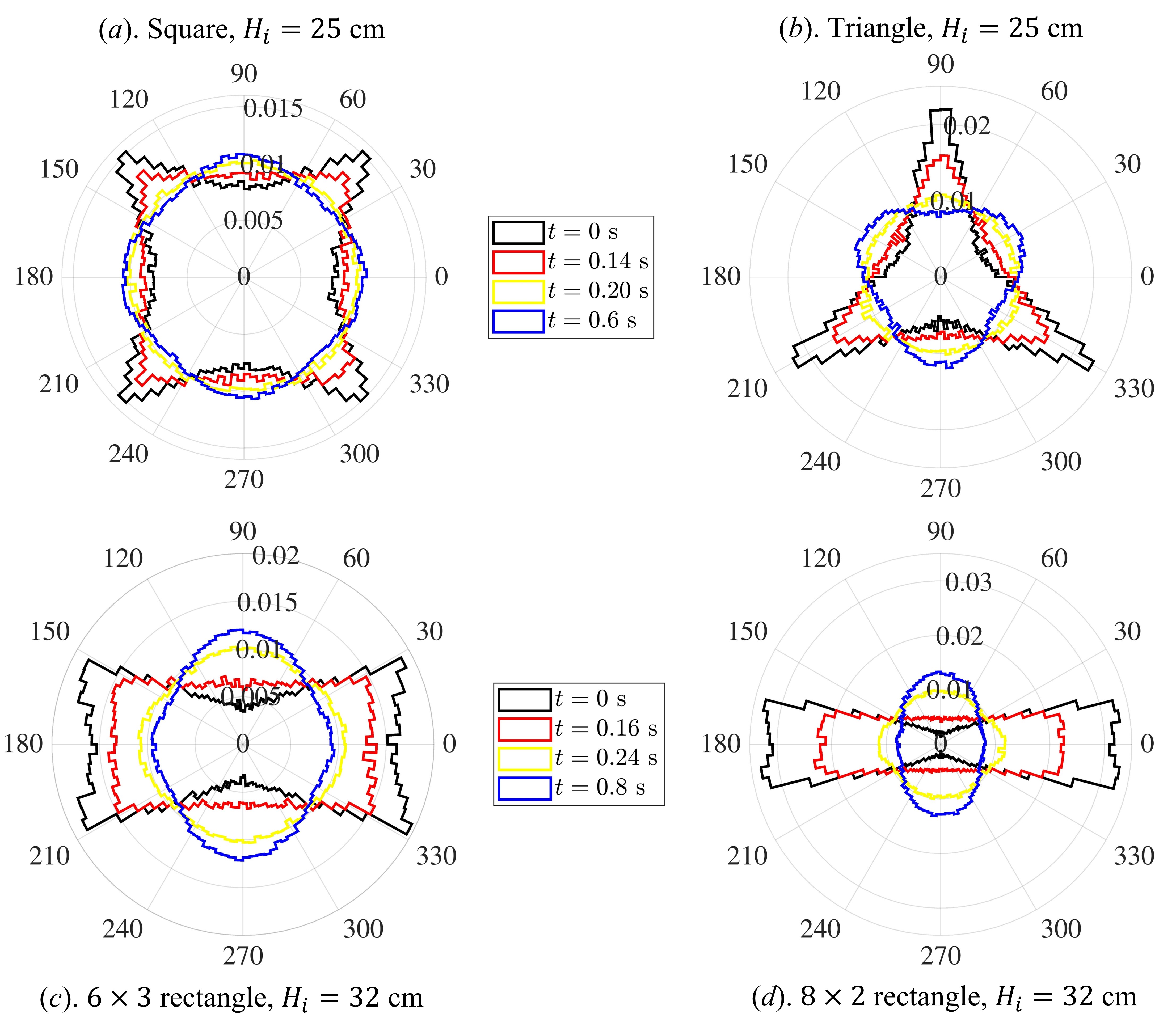}
  \caption{Evolution of polar histograms of tall granular column collapses with respect to time. (a) Square cross-section, $H_i = 25$ cm; (b) triangular cross-section, $H_i = 25$ cm; (c) $6\times3$ rectangular cross-section, $H_i = 32$ cm; and (d) $8\times2$ rectangular cross-section, $H_i = 32$ cm. Sub-figures in the same row share the same legend.}
  \label{fig-histotime}
\end{figure}

Figure \ref{fig-histotime} shows the evolution of polar histograms with respect to time. We only plot the polar histogram of tall columns with four different cross-sections [(a) Square cross-section, $H_i = 25$ cm; (b) triangular cross-section, $H_i = 25$ cm; (c) $6\times3$ rectangular cross-section, $H_i = 32$ cm; and (d) $8\times2$ rectangular cross-section, $H_i = 32$ cm]. Figs. \ref{fig-histotime}(a)-(d) show similar behavior that, when the column is tall enough, the deposition pattern will quickly forget its original shape. The polar histogram quickly develop to be approximately a circle within 0.3 s. However, for tall columns, as we have shown in the previous section, the circular distribution could not last long until it evolves to its final deposition histogram shown in Figs. \ref{fig-square}(d), \ref{fig-triangle}(d), \ref{fig-rect1}(d), and \ref{fig-rect2}(d). This shows that tall granular columns, which can generate tremendous inertia, often collapse in a way that most particles localize into certain directions, and those localized directions are often not the initial dominant direction in original configurations.

\begin{figure}[!ht]
  \centering
  \includegraphics[scale = 0.4]{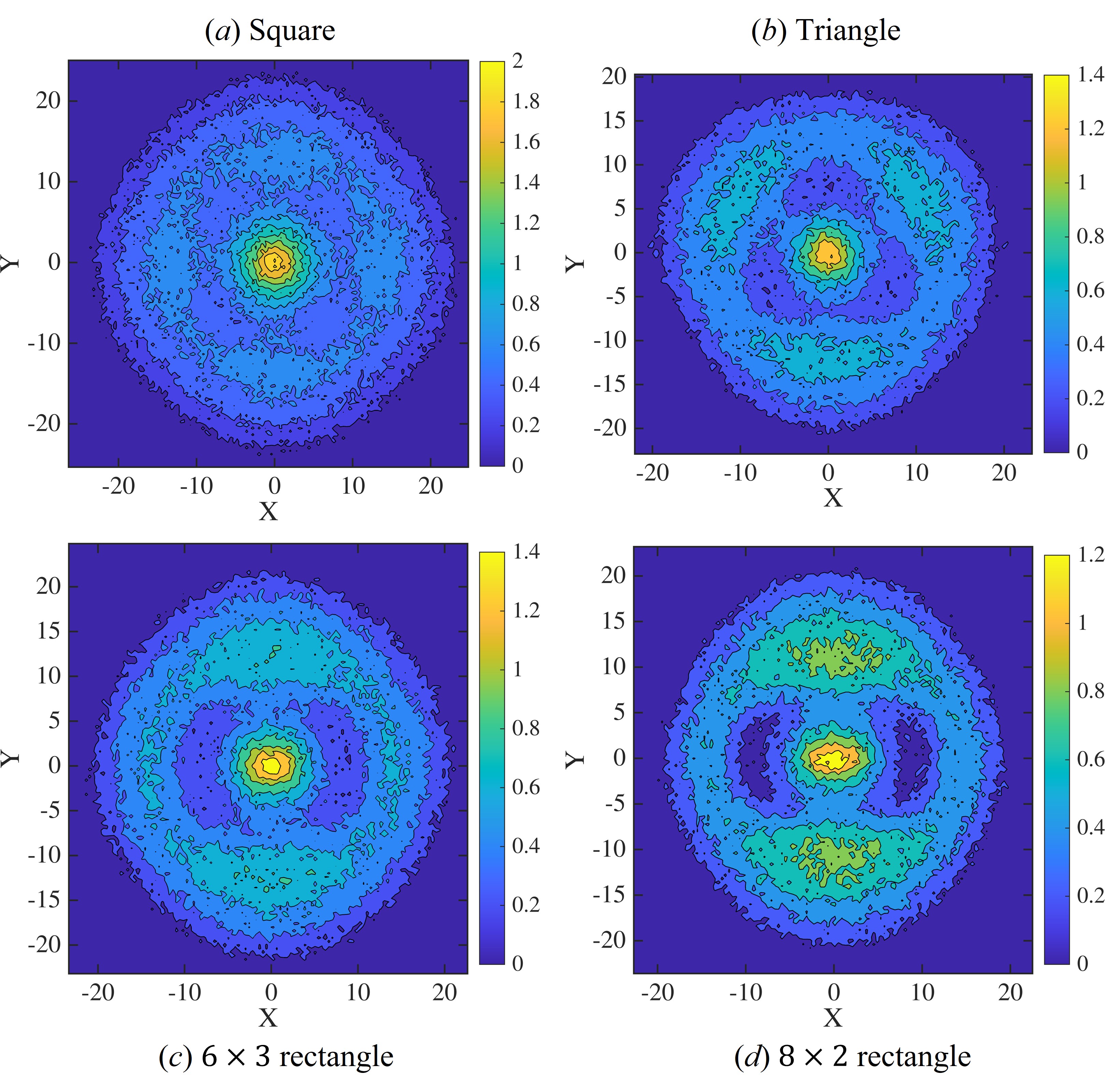}
  \caption{Contour plots of the final deposition of granular columns collapses with four different initial cross-sections: (a) a column with square cross-section and $H_i = 30$ cm; (b) a column with triangular cross-section and $H_i = 30$ cm; (c) a column with $6\times 3$ rectangular cross-section and $H_i = 32$ cm; (d) a column with $8 \times 2$ rectangular cross-section and $H_i = 32$ cm.}
  \label{fig-contour}
\end{figure}

We note that polar histograms may exaggerate the percentage of particles presented in certain directions. Thus, we also plot contour plots of the deposition, in Fig. \ref{fig-contour}, for tall granular column collapses with four different cross-sections. We can see from Fig. \ref{fig-contour}(a) and (b) that particles initially presented in the edge direction tend to travel farther, and more particles seem to rest in the edge direction in the end. The accumulation of particles in the short-edge direction is more obvious for systems with rectangular cross-sections shown in Fig. \ref{fig-contour}(c) and (d), and particles in short edge direction have larger deposition radius. The final deposition pattern is not so exaggerate as the polar histograms, but it still reflects the behavior shown in Figs. \ref{fig-square}(e \& f), \ref{fig-triangle}(e \& f), \ref{fig-rect1}(e \& f), and \ref{fig-rect2}(e \& f), and does not weaken our argument that particles in certain directions prefer to travel farther than other directions, and the classical method for quantifying the run-out behavior is not enough to describe systems with non-circular cross-sections. The detailed deposition pattern of systems with different initial cross-sections is also intriguing and is worth of further explorations in future studies.

\subsection{Analysis with equivalent column radius}
\label{sec-equivR}

\begin{figure}[!ht]
  \centering
  \includegraphics[scale = 0.4]{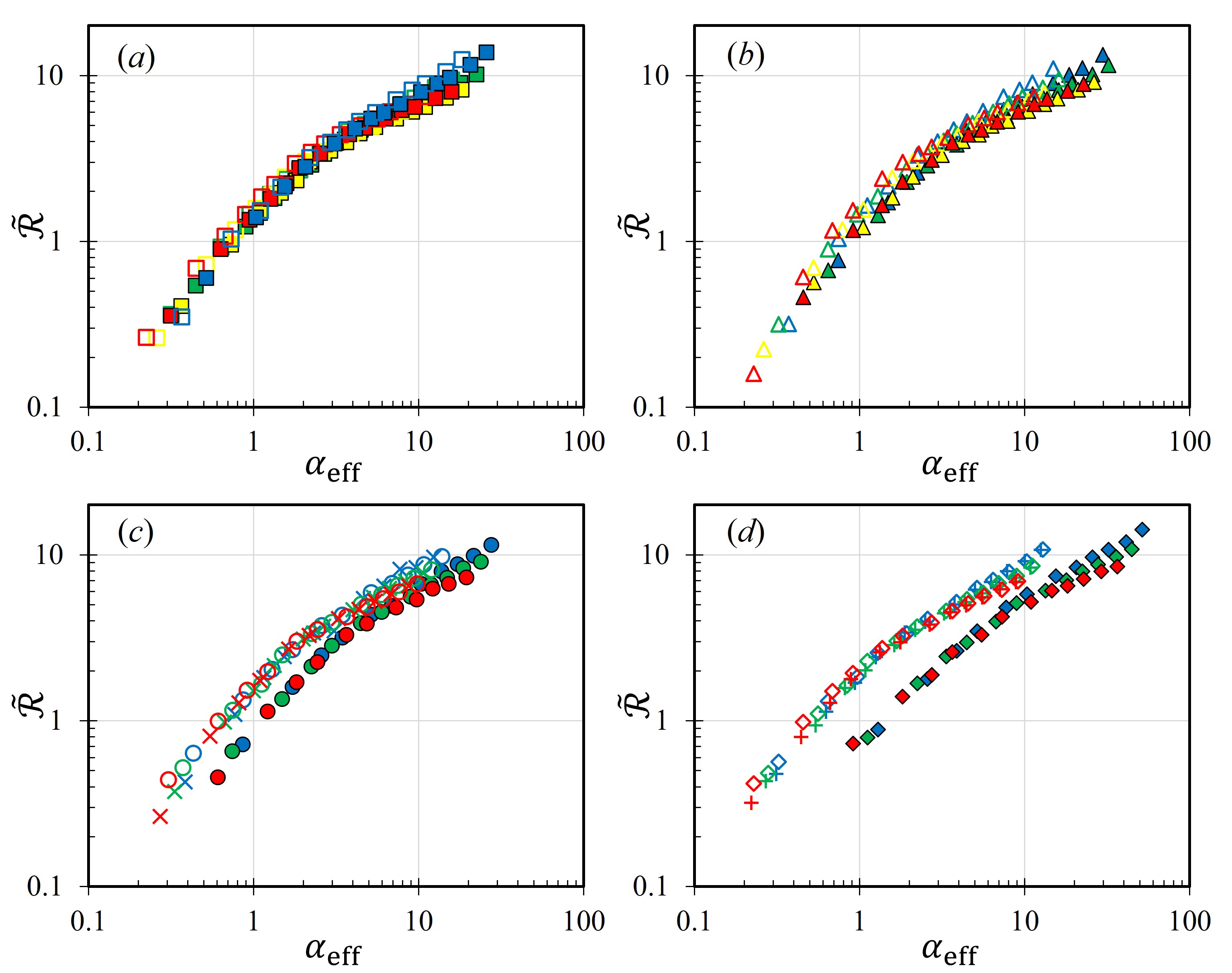}
  \caption{Relationship between the equivalent normalized run-out distance, $\tilde{\mathcal{R}}$, and the effective aspect ratio, $\alpha_{\textrm{eff}}$, of granular columns with (a) square cross-sections, (b) equilateral triangular cross-sections, (c) $6\times3$ rectangular cross-sections, and (d) $8\times2$ rectangular cross-sections. The markers are the same as those in Figs. \ref{fig-square}-\ref{fig-rect2}.}
  \label{fig-equiv}
\end{figure}

In the previous section, we argue that it is not appropriate to use the initial column radius in certain directions to normalize the run-out distance in that direction. In this section, we conveniently suggest an equivalent column radius, the same for any direction of a column, to be defined to obtain a so-called equivalent normalized run-out distance, $\tilde{\mathcal{R}}$, defined by
\begin{equation}
\begin{split}
\tilde{\mathcal{R}} = (R_\infty - R_i)/R_{\textrm{equiv}},\ \ R_{\textrm{equiv}} = \sqrt{A_c/\pi},
\end{split}
\end{equation}
where $R_{\textrm{equiv}}$ is the equivalent column radius and $A_c$ is the area of the cross-section. The advantage of using such a definition of $R_{\textrm{equiv}}$ is that, on one hand, the area-equivalent radius can, in a sense, help us compare granular columns with various cross-sections to axisymmetric columns. On the other hand, when the cross-section is circular, $R_{\textrm{equiv}}$ becomes the same as $R_i$, which makes it convenient to formulate a universal equation to describe the behavior of granular column collapses with any types of cross-section.

Figure \ref{fig-equiv} shows the relationship between $\tilde{\mathcal{R}}$ and $\alpha_{\textrm{eff}}$ of simulations with different cross-sections. In Fig. \ref{fig-equiv}(a), we plot the results of columns with square cross-sections. After changing the $y-$axis to $\tilde{\mathcal{R}}$, all the simulation data collapse nicely onto one curve. The relationship also performs well in terms of granular columns with equilateral triangular cross-sections. However, one thing that concerns us is that the $\tilde{\mathcal{R}}^v - \alpha_{\textrm{eff}}^v$ relationship is slightly above the $\tilde{\mathcal{R}}^e - \alpha_{\textrm{eff}}^e$ relationship, which indicates that, in terms of the equivalent run-out distance, particles in the vertex direction may travel longer equivalent distances than those in the edge direction. Our concern is confirmed by the results of granular column collapses with rectangular cross-sections [Fig. \ref{fig-equiv}(c) and (d)], where, even though there is almost no difference between the $\tilde{\mathcal{R}}^v - \alpha_{\textrm{eff}}^v$ relationship and the $\tilde{\mathcal{R}}^{el} - \alpha_{\textrm{eff}}^{el}$ relationship, the $\tilde{\mathcal{R}}^{es} - \alpha_{\textrm{eff}}^{es}$ relationship is below the other two relationships, especially for columns with $8\times2$ rectangular cross-sections. This can be seen as a failure of implementing the equivalent column radius, yet also an opportunity to perform the finite-size scaling to our work. Similar to what we obtained in our previous work \cite{man2021finitesize}, larger initial radii in vertex and long-edge directions lead to larger relative system sizes, $R_i/d$, and larger normalized run-out distances (in this case, it is the equivalent normalized run-out distance, $\tilde{\mathcal{R}}$). We should also be careful about $R_{equiv}$ that the currently definition of the equivalent cross-section radius, $R_{equiv}$, is rather geometrical than physical. In future works, we should try to link this geometric parameter to the dynamics of particle in different direction.

\subsection{Finite-size analysis}
\label{sec-fss}

\begin{figure}[!ht]
  \centering
  \includegraphics[scale = 0.45]{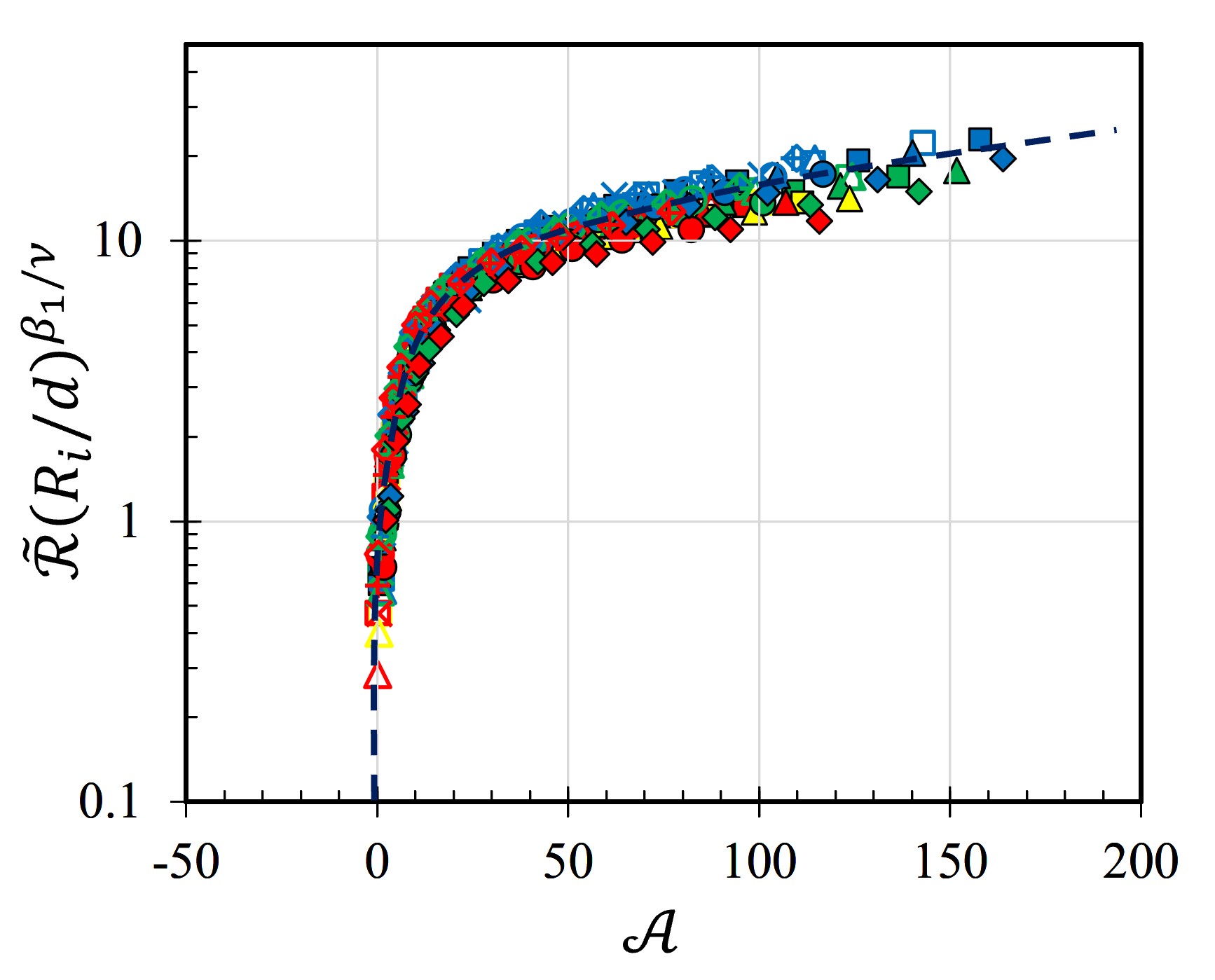}
  \caption{Relationship between $\tilde{\mathcal{R}}(R_i/d)^{\beta_1/\nu}$ and $\mathcal{A}=(\alpha_{\textrm{eff}}-\alpha_{c\infty})(R_i/d)^{1/\nu}$, which considers the size effect of granular columns with different cross-sections. The markers are the same as those in Figs. \ref{fig-square}-\ref{fig-rect2} and the dashed curve is a fitting line}
  \label{fig-fss}
\end{figure}

We summarized the finite-size scaling of axisymmetric granular column collapses as Eq. \ref{eq_fss} in Section \ref{sec:intro}. In our previous work, we analyzed the size effect of granular column collapses\cite{man2021finitesize}. The relative system size, $R_i/d$, ranged from 2.0 to 30.0. We found that, as we increase the relative system size, both the transitional $\alpha$ and the transitional $\mathcal{R}$ decreases in a power-law manner, which led us to perform a finite-size scaling with results shown in Eq. \ref{eq_fss}. We also confirmed the size effect by measuring the relative correlation length scale, which also showed a power-law decay with respect to $\alpha_{\textrm{eff}}$ and decreased as we increased the system size. In this study, the normalized run-out distance is calculated using the equivalent initial radius in each direction. We plot the relationship between $\tilde{\mathcal{R}}(R_i/d)^{\beta_1/\nu}$ and $\mathcal{A}=(\alpha_{\textrm{eff}}-\alpha_{c\infty})(R_i/d)^{1/\nu}$ of all the simulation results in Fig. \ref{fig-fss} with the same scaling parameters as in Ref. \cite{man2021finitesize} that $\nu \approx 1.39$ and $\beta_1 \approx 0.28$. The following relationship is expected
\begin{equation} \label{eq_fss2}
    \begin{split}
        \tilde{\mathcal{R}} = \left({R_i^{\theta}}/{d}\right)^{-\beta_1/\nu}\mathcal{F}_r\left[(\alpha_{\textrm{eff}}^{\theta}-\alpha_{c\infty})\left({R_i^{\theta}}/{d}\right)^{1/\nu}\right]\ ,
    \end{split}
\end{equation}
where the superscript $\theta$ denotes the direction in which we take the measurement, and can be replaced with $e$, $v$, $el$, or $es$. The results in Fig. \ref{fig-fss} agree with our expectation that all the simulation data of $\tilde{\mathcal{R}}(R_i/d)^{\beta_1/\nu} - \mathcal{A}$ relationship with different cross-sections and different measuring directions  form a function, $\mathcal{F}_r(\cdot)$.

The collapse of the $\tilde{\mathcal{R}}(R_i/d)^{\beta_1/\nu} - \mathcal{A}$ relationship indicates that the finite-size scaling is still valid when we consider a non-circular cross-section. In other works related to the size effect of granular column collapses, sizes effects refer to different sizes of different columns. This work shows that, even within one granular column, size effect should still be considered. We realize that, without a proper definition of $\mathcal{F}_r(\cdot)$, it is impossible to accurately calculate the run-out distance with any cross-section. Nevertheless, this work helps formulate a universal solution to describe the run-out behavior considering frictional coefficients, initial conditions, and geometric factors, which is convenient to utilize when evaluating some geo-hazards, such as landslides and pyroclastic flows, where both the system size and the geometric factor are important for understanding the mobility of them. We note that, the analysis, based on equivalent cross-section radii, may fail to work for systems with extremely slender cross-sections. A more proper definition of $R_{equiv}$ need to be discussed with a clearer physical definition in future works.

\section{Experimental Evidence of the Cross-section influence}
\label{sec:exp}

\begin{figure}[!ht]
  \centering
  \includegraphics[scale = 0.35]{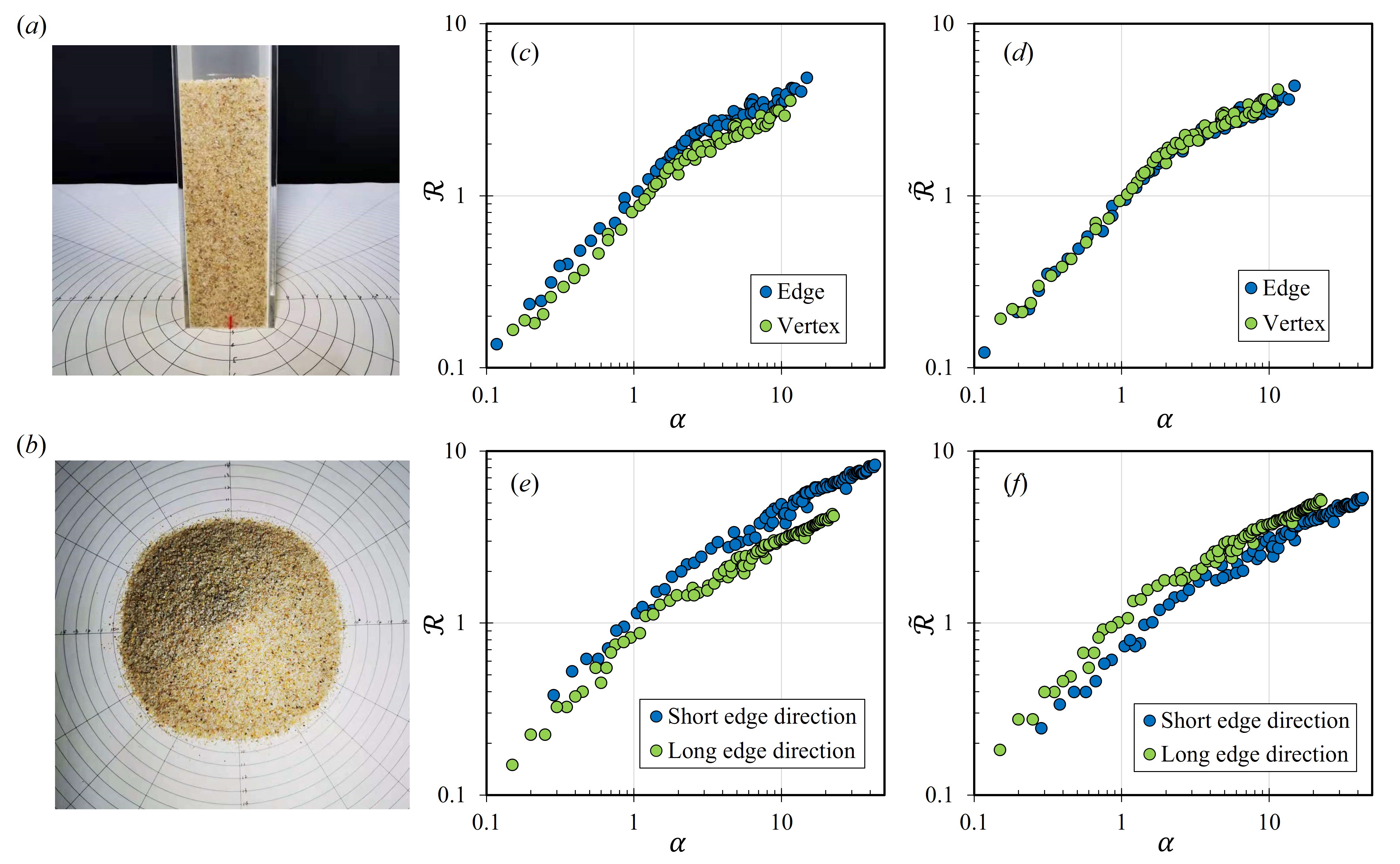}
  \caption{(a) and (b) shows the initial condition and the final granular pile of a granular column collapse from a tube with a square cross-section; (c) and (d) show the experimental results of granular columns with 50 mm $\times$ 50 mm square cross-sections; and (e) and (f) show experimental results of granular columns with 40 mm $\times$ 21 mm rectangular cross-sections.}
  \label{fig-exp}
\end{figure}

To show the evidence of the influence of changing cross-section shapes, we performed two sets of experiments of granular column collapses from both square cross-section and rectangular cross-section tubes with sand particles of diameters ranging from 0.1 mm to 0.2 mm. The aim of this section is showing the orientation anisotropy presented in granular column collapses with non-circular cross-section, instead of presenting a comparison between simulations and experiments. We considered two different types of granular columns: (1) granular columns with 50 mm $\times$ 50 mm square cross-sections; and (2) granular columns with 40 mm $\times$ 21 mm rectangular cross-sections. For both experimental setups, after placing sand particles into the plastic tube, we measured the initial height of the granular packing, $H_i$. Sand particles are dropped in from the top of the tube so that the initial condition resembles a randomly loose packing of granular system. Then, the tube was manually lifted to release all the particles to form a granular pile. For columns with square cross-sections, we measured run-out distances in both vertex and edge directions. For column with rectangular cross-sections, we only measured run-out distances in both long edge and short edge directions as previously defined in Fig. \ref{fig-setup}(d) and (e). The corresponding normalized run-out distances, $\mathcal{R}$ and $\tilde{\mathcal{R}}$, can be obtained accordingly. Since the frictional coefficient was kept as constant during experiments, we use the initial aspect ratio, $\alpha$, as the $x-$axis in the figures (Fig. \ref{fig-exp}), without considering the corresponding effective aspect ratio, $\alpha_{\textrm{eff}}$.

Fig. \ref{fig-exp}(a) and (b) shows the initial condition and the final granular pile of a granular column collapse from a tube with a square cross-section. We can see that, even thought the final deposition pattern still shows some clue of the initial square cross-section, Particles in the edge direction are catching up to make the final pattern approximately circular. Figure \ref{fig-exp}(c) shows the relationship between the normalized run-out distance and the initial aspect ratio of granular columns with square cross-sections. For the same initial aspect ratio, the normalized run-out distance in the edge directions is apparently larger than that in the vertex directions which shows similar behavior to that in Fig. \ref{fig-square}(e). As we convert the $y-$axis into the equivalent normalized run-out distance, $\tilde{\mathcal{R}}$ in Fig. \ref{fig-exp}(d), the data collapse onto one curve, which is the same as that in Fig. \ref{fig-equiv}(a). No obvious size effect is shown in Fig. \ref{fig-exp}(d) because that $R_i^v$ is only approximately 1.4 times of $R_i^e$. 

Figure \ref{fig-exp}(e) shows experimental results of granular columns with 40 mm $\times$ 21 mm rectangular cross-sections. Similar to Fig. \ref{fig-rect1}(e), the normalized run-out distance in the long-edge direction is larger than that in the short-edge direction. As we shift the measuring direction from the edge direction to the vertex direction, the transitional initial aspect ratio also changes accordingly. In Fig. \ref{fig-exp}(f), we plot the relationship between $\tilde{\mathcal{R}}$ and $\alpha$ in both short-edge and long-edge directions. The equivalent normalized run-out distance in the long-edge direction becomes larger than that in the short-edge direction, and significant size effect can be observed since the initial radius in the long-edge direction is almost twice the initial radius in the short-edge direction. The experimental work shows significant similarities in both cross-section shape influence and size effects with the simulation results we presented earlier. This set of experiments confirms that, due to the influence of cross-sesction anisotropy, particles tend to flow towards certain directions, and size effect caused by different initial radii for granular columns with non-circular cross-sections should be considered.

\section{Conclusions}
\label{sec:conclude}

In this paper, we explore the influence of cross-section shapes on the collapse of granular columns. Four different cross-sections are considered: square, equilateral triangular, $6\times3$ rectangular, and $8\times2$ rectangular. We show that the normalized run-out distance in the edge direction tends to be larger than that in the vertex directions when they have similar initial aspect ratios or effective aspect ratios. It indicates that the geometric factor plays an important role in determining the run-out distance of granular column collapses. This might be important when dealing with other problems in granular physics. 

We propose that using an equivalent initial radius, instead of using the initial radius in the measuring direction, to normalize the run-out distance in that direction, performs better in terms of collapsing the simulation data. In this study, we defined the equivalent initial radius as $R_{\textrm{equiv}}=\sqrt{A_c/\pi}$, which is the radius of the circle with the same area as the cross-section. Using the equivalent radius, we could obtain the equivalent normalized run-out distance, $\tilde{\mathcal{R}}=(R_{\infty}-R_i)/R_{\textrm{equiv}}$ in any measuring directions. However, the $\tilde{\mathcal{R}} - \alpha_{\textrm{eff}}$ relationship fails when considering the columns with rectangular cross-sections, especially when the length of the cross-section is much larger than the width. This leads us to perform the finite-size analysis to obtain the $\tilde{\mathcal{R}} - \alpha_{\textrm{eff}}$ relationship, and results in a universal relationship which could simultaneously consider the influence of inter-particle friction, particle/boundary friction, initial aspect ratio, and geometric factors.

We conclude that size effect should be considered not only when dealing with different granular columns but also when calculating the run-out distance in different direction within the same granular column. This becomes important when we have a geo-hazard source with an irregular initial geometry. We also note that, due to the nature of Voronoi diagrams, the initial packing of the system is denser and more ordered than natural granular systems. We also believe that the run-out behavior of granular columns should be linked to the rheological properties of granular systems, which implies that the rheology of granular systems with different system sizes and different frictional coefficients should be included in the $\mathcal{F}_r[\cdot]$ function so that the scaling law could become more physics-based. In fact, the body of study we have introduced here offer clues on the true form of the rheology law governing this behavior. Furthermore, a similar critical point should be observed at some micro-mechanical quantity and should be included in the true rheological law. Further investigations to link the behavior of idealized granular system and realistic geophysical flows  and include granular rheology into the analysis of granular column collapses are still needed, and will be presented in future publications.

\section*{Acknowledgements}
The authors acknowledge the financial support from the General Program (NO. 12172305) of the National Natural Science Foundation of China and Westlake University, and thank the Westlake University Supercomputer Center for computational resources and related assistance. The authors would like to thank Prof. Ling Li for continuing helpful discussions related to this study. All the simulations were conducted using MechSys (\url{mechsys.nongnu.org})\cite{Mechsys}.

\section*{Conflict of interest}
The authors declare that they have no conflict of interest.




 \bibliographystyle{elsarticle-num} 
 \bibliography{CrossSectionEffect}





\end{document}